\documentclass[aps,prd,superscriptaddress,showpacs,preprintnumbers,floatfix]{revtex4}
\usepackage{graphicx,color}
\usepackage{amsmath}
\usepackage{amsfonts}
\usepackage{amssymb}
\newcommand{\mup}{\kappa}
\begin{document}
\title{Phenomenology of TMDs}
\author{S.~Melis}
\affiliation{Dipartimento di Fisica, Universit\`a di Torino,
             Via P.~Giuria 1, I-10125 Torino, Italy}
%
%



 \begin{abstract}         
 We present a review of current Transverse Momentum Dependent (TMD) phenomenology
 focusing our attention on the unpolarized TMD parton distribution function and the Sivers function.
 The paper introduces  and comments about the new Collins-Soper-Sterman (CSS) TMD evolution formalism
 .
 We make use of a selection of results obtained by several groups to illustrate
 the achievements and the failures of the simple Gaussian approach
 and the TMD CSS evolution formalism. 
\end{abstract}
\maketitle
\section{Introduction}
\label{intro}
In the parton model a fast moving nucleon can be thought as a collection of quasi free partons: quarks and gluons.
Their number density distribution is called Parton Distribution Function (PDF).
As we just mentioned, partons are \textit{quasi} free. In fact, they ``strongly'' interact continuously.
The PDFs cannot be calculated \textit{ab initio} using perturbative QCD,
they are quantities that must be extracted from the experimental data.
This can be done thanks to factorization theorems that are valid only in certain kinematical regions.
They allow to write a (measurable) hadronic cross section
as the convolution of a perturbative calculable partonic cross section with the (non-perturbative) PDFs (and/or fragmentation functions as appropriate).
Then the property of universality tells us how to connect the PDFs measured
in a rather different variety of scattering processes involving nucleons.
For ordinary unpolarized PDFs this connection is trivial, since the PDFs are the same for any factorizable process.
Universality and factorization clearly convert the PDF in the quantity that describes the ``nucleon structure'' at high energies.

If we consider only the longitudinal degrees of freedom of the partons, our PDF is called ``collinear''.
The collinear PDFs, $f(x)$, can be interpreted as the number density of partons carrying a light-cone fraction $x$ of the nucleon momentum.
If, instead, we consider in addition the transverse component of the parton motion,
we speak about Transverse Momentum Dependent PDFs (TMD PDFs or simply TMDs), $f(x,\mathbf{k}_{\perp})$.
Notice that the parton motion is an internal degree of freedom of the nucleon
and therefore not all the observables are sensitive to it.
For instance, total cross sections do not give any direct information on the nucleon internal structure, i.e. on the PDFs.
Similarly, if we consider inclusive processes like the Deep Inelastic Scattering ($l p\to l^{\prime}+X$),
we are only sensitive to the longitudinal degrees of freedom (the Bijorken $x_B$).
In this case we can get information on the collinear PDFs but not on the TMDs.
To access a fully three-dimensional TMD we need a process like the Semi Inclusive DIS ($l p\to l^{\prime} h+X$)
where we can observe the transverse momentum of the produced final hadron.

In Semi Inclusive DIS (SIDIS) processes a second source of transverse momenta is given of course by the Fragmentation Functions (FFs).
FFs describe the hadronization process
of the partons and, similarly to the PDF case, we can have collinear or TMD FFs.

Further (interesting) complications arise if we consider also the nucleon polarization as a new degree of freedom.
In this case we can define two additional collinear PDFs (transversity and helicity functions) and eight TMDs in total.

Collinear PDFs (and FFs) have been extensively studied in literature both experimentally and theoretically.
TMDs instead are more subtle objects and only recently theoretically based definitions have been established.
Some of the TMDs exhibit non-trivial universality properties.
It is the case of the Sivers function, which changes sign when observed in SIDIS rather than in Drell-Yan (DY) processes.
These non-trivial properties make them an exceptional laboratory to study QCD.
For a recent general introduction on TMDs see for example Ref.~\cite{Barone:2010zz}.

Presently, our main sources of information on the TMDs are contained in SIDIS and Drell-Yan data.
In fact, for these two processes the TMD factorization is well established, see for instance Ref.~\cite{Collins:2011zzd}.
As said before, if we want to study the TMDs we have to restrict ourselves to those observable which are sensitive
to the transverse momentum distribution of the quarks and the gluons.
Mainly, for DY processes, these are the absolute cross sections as function
of transverse momentum of the dilepton pair ({\it i.e.} of the virtual photon $\gamma^*$), $P_T$,
or the average transverse momentum $\langle P_T^2\rangle$ or the azimuthal asymmetries.
Similarly in SIDIS we can consider the multiplicity data,
the average transverse momentum of the produced hadron or, again, the azimuthal asymmetries.
These two processes involve convolutions of two TMD PDFs (DY) or of one TMD PDFs and one TMD FF (SIDIS).
They give us complementary information on the TMDs.
A complete picture would require also data from transverse momentum dependent observables in $e^+e^-$ scattering processes,
which, unfortunately, are not available yet.

A global analysis of DY and SIDIS data is the main target of the present phenomenological investigation on the TMDs.
This kind of analysis is actually rather difficult
because of the fragmentary pieces of (experimental) information
that  we have at our disposal.
One of the difficulties is that present (transverse momentum dependent)
SIDIS and Drell-Yan data span very different regions
in the center of mass energies, $\sqrt{s}$, transfer momenta, $Q^2$, and transverse momenta, $P_T$.
In particular DY data cover a very large region in $\sqrt{s}$, from tens of GeVs up to tens of TeVs,
a large $Q$-region from the $J/\Psi$ resonance up to the mass of the $Z_0$ boson.
The corresponding $P_T$ then runs from hundreds of MeVs for the low energy DY experiments up to hundreds of GeVs for the recent Tevatron and LHC data.
On the contrary, transverse momentum dependent data from SIDIS come mainly from JLAB, HERMES and COMPASS experiments, which run at very low $\sqrt{s}$ (
from 3.6 up to 18 GeV) covering a small region in $Q$ (from 1 up to 3.2 GeV mainly) and with small $P_T$ ($0.1<P_T\lesssim 1.5$ GeV roughly).
We can see that current SIDIS data are at the limit of applicability of perturbative QCD.
Adding to that the fact that we do not have independent information on fragmentation from electron-positron annihilation processes,
it is clear that a global fit of DY and SIDIS data is a rather difficult task.

\section{Unpolarized Drell-Yan data phenomenology}
\label{Sec:Drell-Yan}
\subsection{Gaussian models}
\label{SubSec:Drell-Yan-Gauss}
\begin{figure}
\centering
\includegraphics[width=11cm,clip]{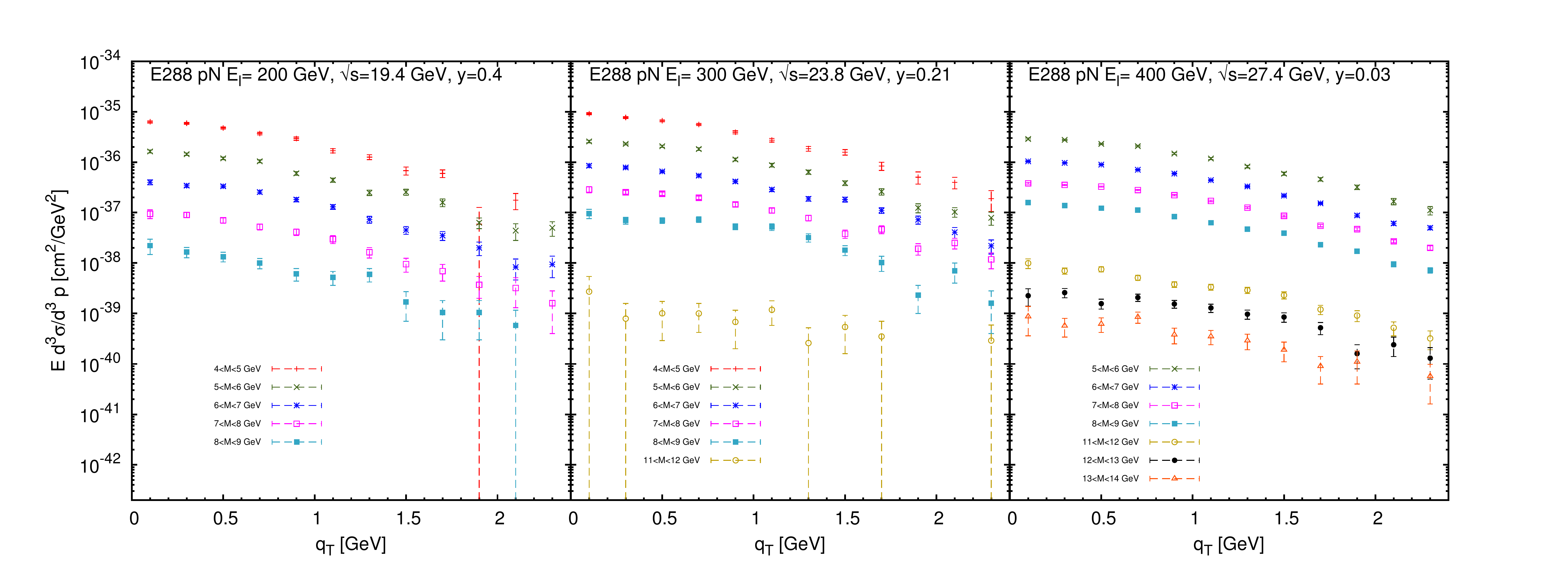}
\caption{Data for the Drell-Yan invariant cross section as function of $P_T$ collected at FERMILAB, E288 experiment~\cite{Ito:1980ev},
at three different energies of the beam:
200 GeV (left panel), 300 GeV (central panel) and 400 GeV (right panel)}
\label{fig-1}       
\end{figure}
The transverse momentum spectra of low energy Drell-Yan are quite peculiar. At low $P_T$ in fact, the spectra seem to have a Gaussian shape.
See, for instance, Fig.~\ref{fig-1}.
To test this hypothesis one can use a simple Gaussian model for the TMD PDF:
\begin{equation}
 {f}_{q/p}(x, k_{\perp};Q)=f_{q/p}(x;Q) \,\frac{e^{-k_\perp^2/\langle k_{\perp}^2\rangle}}
{\pi\langle k_{\perp}^2\rangle}
\label{GaussianPDF}
\end{equation}
where $f_{q/p}(x;Q^2)$ is the usual collinear PDF.
Notice that in this simple model the $x$ and $k_{\perp}$ dependencies are factorized and the $Q^2$ dependence is only in the collinear PDF.
Inserting this model in the Drell-Yan Born cross section we get:
\begin{equation}
\frac{d\sigma}{d P_T^2}\propto \frac{\alpha_{em}}{M^2}\sum_q f_{q/h_1}(x_1)\bar{f}_{q/h_2}(x_2)\,\frac{\exp(-P_T^2/\langle P_T^2\rangle)}{\pi\langle P_T^2\rangle}\,.
\label{xsecg}
\end{equation}
The final distribution in $P_T$ is Gaussian and its width is given by the sum of the transverse momenta of the quark/antiquark pair:
$\langle P_T^2\rangle=\langle k_{\perp 1}^2\rangle+\langle k_{\perp 2}^2\rangle$.
Eqs.~\ref{GaussianPDF} and~\ref{xsecg} hold under the assumption that the transverse momentum distribution is the same for quark and antiquark, for any flavor.
If we further consider a nucleon-nucleon scattering the model simplifies considerably
having only one parameter:
\begin{equation}
\langle P_T^2\rangle=2\langle k_{\perp}^2\rangle.
\end{equation}

\begin{figure}
\centering
\includegraphics[width=5.7cm,clip]{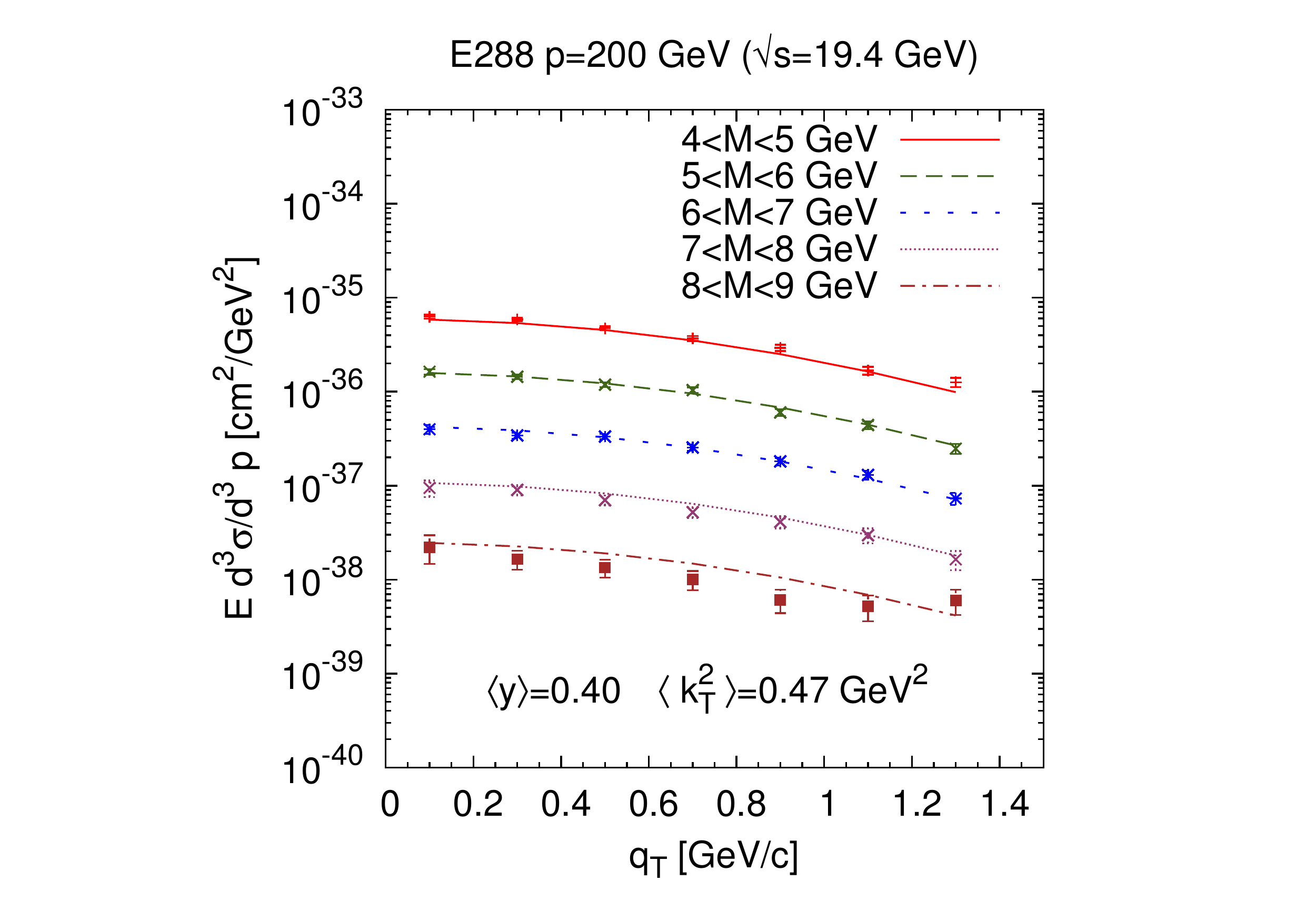}
\hspace{-1.65cm}
\includegraphics[width=5.7cm,clip]{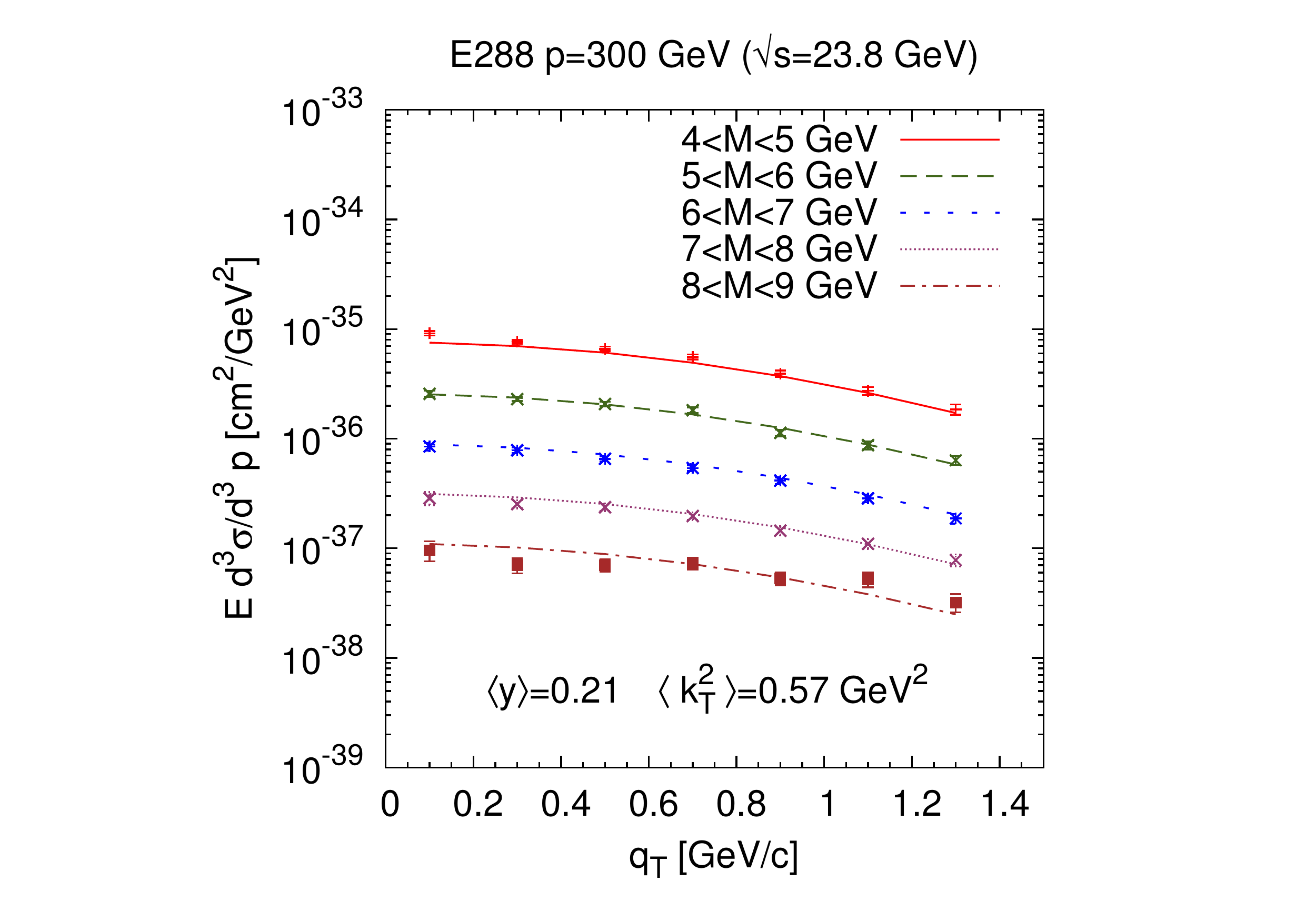}
\hspace{-1.65cm}
\includegraphics[width=5.7cm,clip]{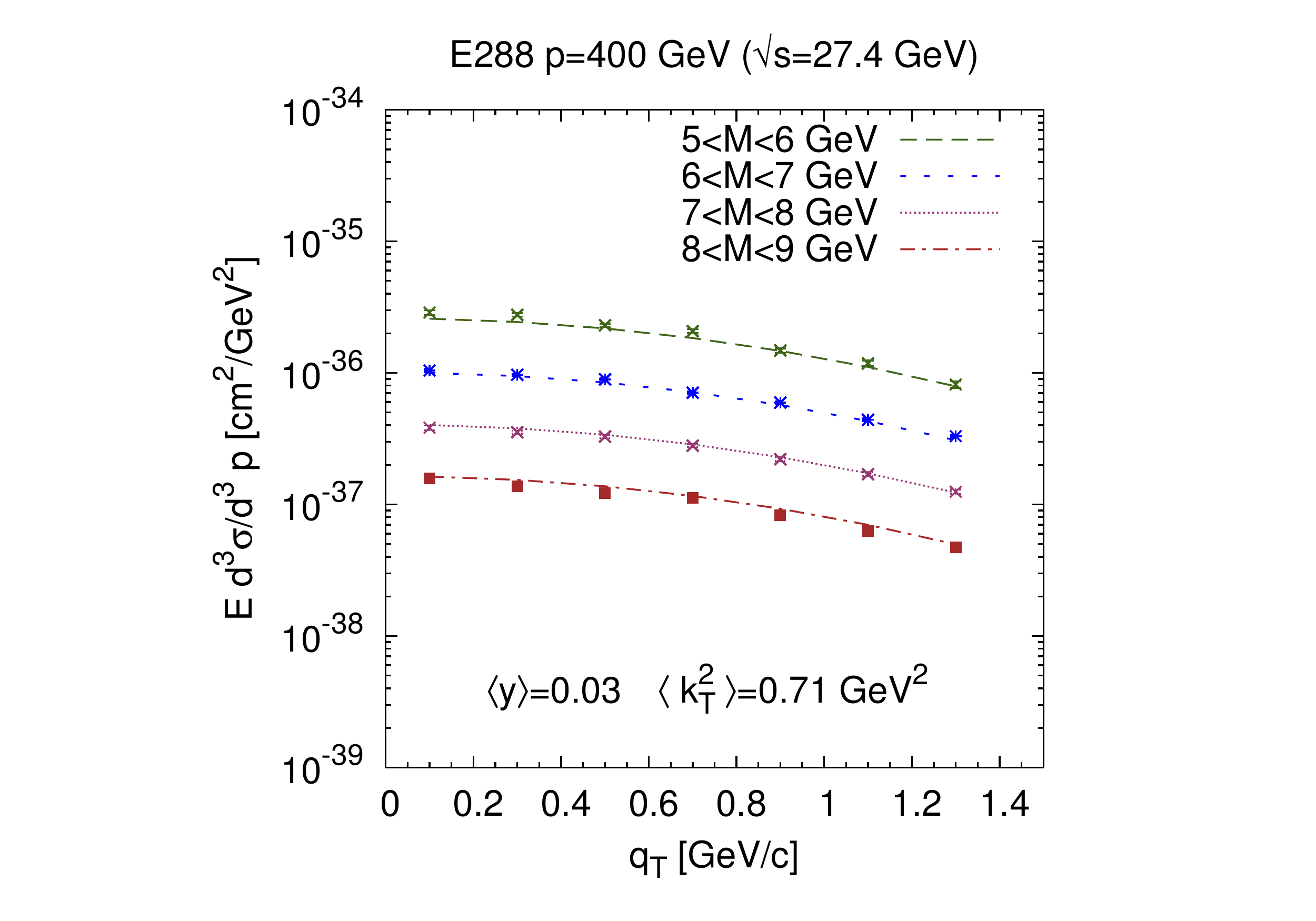}
\caption{Gaussian fit of E288 data~\cite{Ito:1980ev} using the model in Eq.~\ref{GaussianPDF}. Each set of data at different $s$ is fitted separately.}
\label{E288fit}       
\end{figure}

\begin{figure}
\centering
\includegraphics[width=5.7cm,clip]{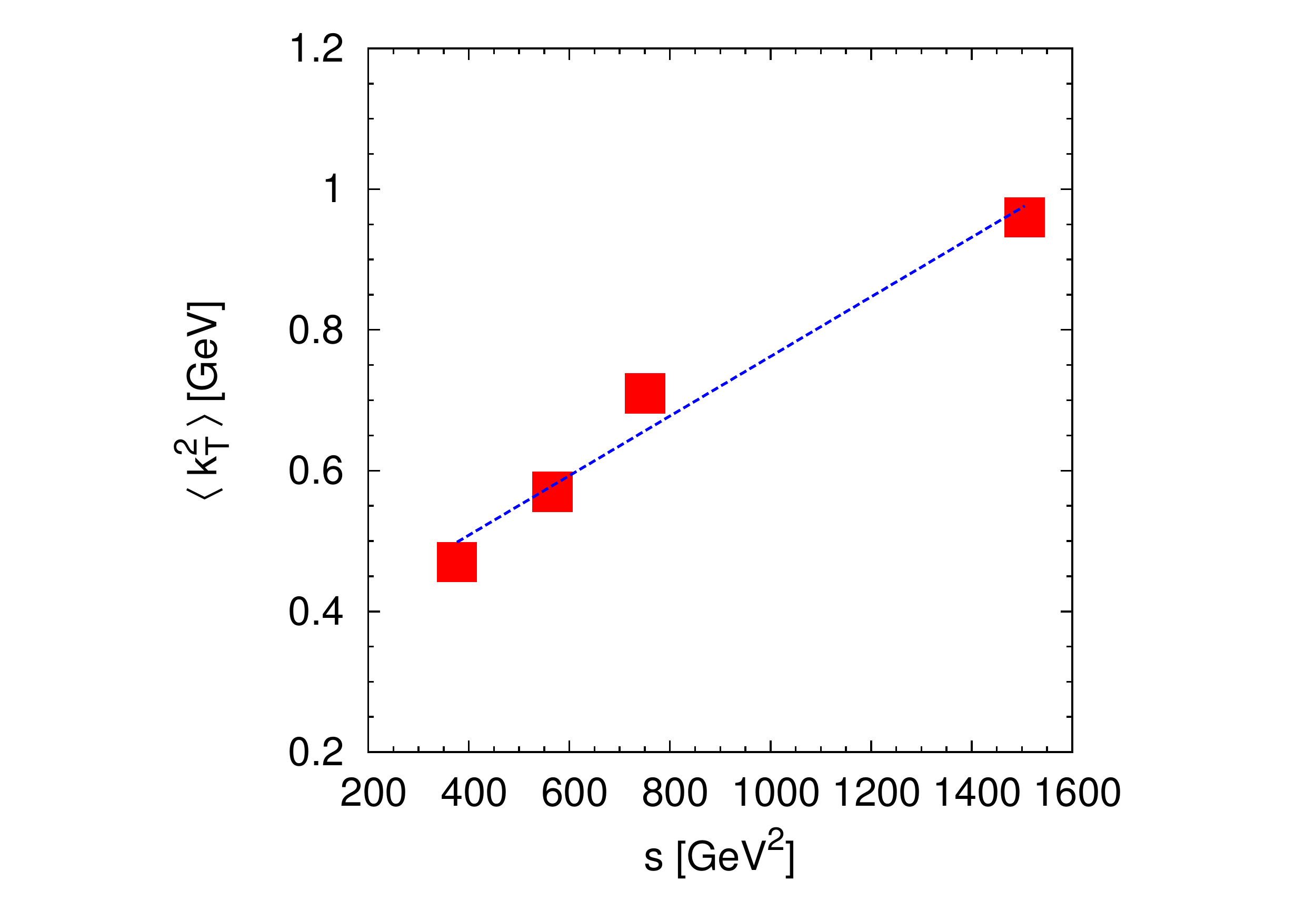}
\includegraphics[width=5.7cm,clip]{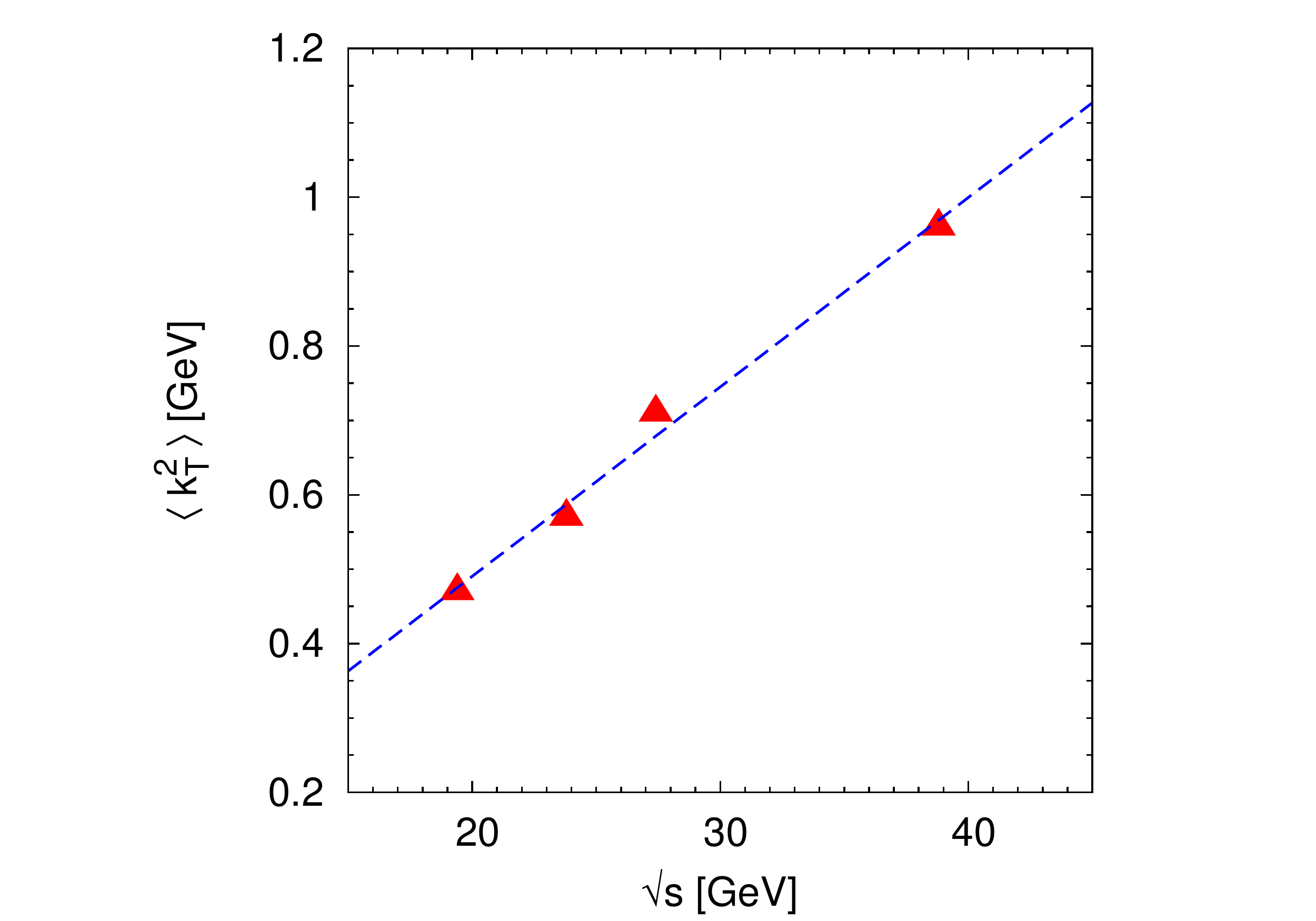}
\caption{$\langle k_{\perp }^2\rangle$ extracted from FERMILAB E288~\cite{Ito:1980ev} and E605~\cite{Moreno:1990sf} data as function of $s$ (left panel) and $\sqrt{s}$ (right panel)}
\label{ktfit}       
\end{figure}

If we fit the data in the left panel of Fig.~\ref{fig-1}, corresponding to the E288 data at 200 GeV, using this model and allowing for one free normalization parameter,
we discover that the Gaussian model describes the shape of the data perfectly well: see left panel of Fig.~\ref{E288fit}.
Moreover it describes also the $Q^2\equiv M^2$ behavior of the data
which is given manly by the $1/M^2$ dependence of the $q\bar{q}$ cross section and by the DGLAP evolution equation of the collinear PDFs.
If one repeats this exercise for the other two energies of the E288 experiment, one discovers that each experiment can be described by a Gaussian
but the width increases with the energy of the experiment.
The proportionality between $s$ and $\langle k_{\perp }^2\rangle$ is difficult to establish.
One can plot the extracted value of  $\langle k_{\perp }^2\rangle$ versus $s$ or $\sqrt{s}$ as in Fig.~\ref{ktfit}. 
Approximately we can see a linear dependence of $\langle k_{\perp }^2\rangle$ to some power of $s$.
A linear dependence on $s$ is predicted by QCD at large $P_T$~\cite{Altarelli:1977kt}.
A linear dependence on $s$ has been also claimed for DY and SIDIS in Ref.~\cite{Schweitzer:2010tt}. However the linearity is very rough and approximate.
\subsection{Soft gluon resummation}
\label{SubSec:Drell-Yan-Soft}
\begin{figure}
\centering
\includegraphics[width=6cm,clip]{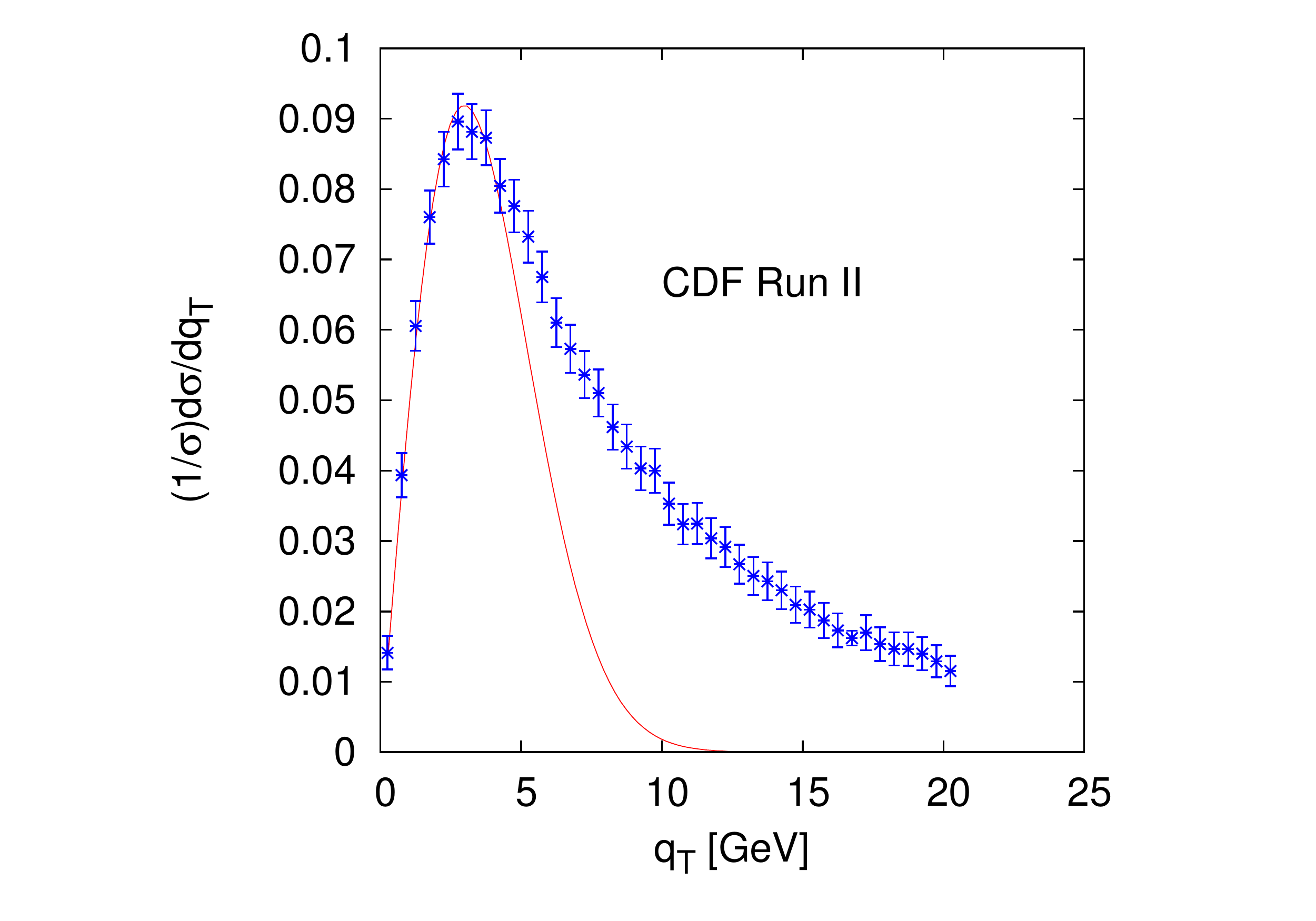}
\caption{CDF Run II data~\cite{Aaltonen:2012fi} and a Gaussian model (red solid line). The Gaussian model is not able to describe the tail of the $P_T$ spectrum.}
\label{tevatron_fit}       
\end{figure}

When considering high energy DY data like Tevatron or LHC data, it is clear that the spectrum is Gaussian only at very small $P_T$
and the previous models cannot describe all the data, as shown, for instance, in Fig.~\ref{tevatron_fit}.
In this case, in fact, the region of the spectrum where $\Lambda_{\textrm{QCD}}\ll P_T\ll Q$ is generated by radiative gluon emissions.
These emissions are partly calculable in QCD using soft gluon resummation techniques and one can write~\cite{Collins:1984kg}:
\begin{equation}
\frac{1}{\sigma_0} \frac{d\sigma}{dQ^2dy d P_T^2}=\int\frac{d^2 \boldsymbol{b}_T e^{i \boldsymbol{P}_T\cdot\boldsymbol{b_T}}}{(2\pi)^{2}}\Bigg\{\sum_j\! e^2_jW_j(x_1,x_2,b_T,Q)\Bigg\}+Y(x_1,x_2,P_T,Q)\label{MasterCSS}
\end{equation}
where the term $Y$ is the part of the collinear cross section which is regular at $q_T\rightarrow 0$, while $W$ is the term that resummes
the radiative gluon contributions. The resummation is performed in the $b_T$ space, the Fourier conjugate space of the transverse momentum space.
The $b_T$ space is preferred in this type of calculation because the momentum conservation can be taken into account quite easily. 
Explicitly the $W$ term reads:
\begin{equation}
W_j(x_1,x_2,b_T,Q)= \exp\left[S_j(b_T,Q)\right]\sum_{i,k} C_{ji}\otimes f_{i}(x_1,C_1^2/b_T^2)\,\, C_{\bar{j}k}\otimes f_{k}(x_2,C_1^2/b_T^2)
\label{W}
\end{equation}
where
\begin{equation}
S_j(b_T,Q)=-\int_{C_1^2/b_T^2}^{Q^2}\frac{d \mup^2}{\mup^2}\left[A_j(\alpha_s(\mup))\ln\left(\frac{Q^2}{\mup^2}\right)+B_j(\alpha_s(\mup))\right] 
\label{S}
\end{equation}
is the so called Sudakov form factor. $A_j$ and $B_j$ are perturbative coefficients that can be calculated in QCD.
Here $C_1=2\exp(-\gamma_E)$ where $\gamma_E$ is the Euler's constant. The subscript $j$ indicates that the coefficients are different for $q\bar{q}$ initiated
processes (like ordinary DY) or $gg$ fusion processes like Higgs bosons production. 
The symbol $\otimes$ in Eq.~\ref{W} represents the usual collinear convolution
between the Wilson coefficients $C_{ji}$ (calculable in QCD) and the collinear PDF $f_i(x,C_1/b_T)$.

Notice that the scale at which the PDF
is evaluated is not $Q^2$ but rather $\mu=C_1/b_T$ and that the integral in the Sudakov form factor runs from $\mu$ to $Q$.
It is clear that at large $b_T$ (i.e. at small $k_{\perp}$) this approach cannot be valid because the scale $\mu$ becomes very small, consequently
we are entering in a non-perturbative regime.
To solve this problem one can adopt a phenomenological approach and define a procedure to freeze the scale $b_T$.
The most common prescription is the so called $b_*$ prescription. 
\begin{equation}
b_*=\frac{b_T}{\sqrt{1+\frac{b_T^2}{b_{max}^{2}}}}\,.
\label{bstar}
\end{equation}
Here $b_{max}$ is a free parameter of the order of 1 GeV. The $b_*$ prescription consists in replacing:
\begin{equation}
b_T\longrightarrow b_*\qquad \mu=C_1/b_T \longrightarrow \mu_b=C_1/b_*
\label{mub_bt_replace}
\end{equation}
 in Eq.~\ref{W} and~\ref{S}, and introducing a non-perturbative, phenomenological function
\begin{equation}
F_{NP}(x_1,x_2,b_T,Q)= \frac{W_j(x_1,x_2,b_T,Q)}{W_j(x_1,x_2,b_*,Q)} 
\end{equation}
to describe the large $b_T$ behavior. The final expression of $W$ then reads:
\begin{equation}
W_j(x_1,x_2,b_T,Q)=\sum_{i,k}\exp\left[S_j(b_*,Q)\right]\Big[C_{ji}\otimes f_{i}\left(x_1,\mu_b\right)\Big]\,\,\Big[C_{\bar{j}k}\otimes f_{k}\left(x_2,\mu_b\right)\Big]
F_{NP}(x_1,x_2,b_T,Q)\,.
\label{W2}
 \end{equation}

Frequently used models for $F_{NP}$ are the Davies-Webber-Stirling (DWS), Landinsky-Yuan (LY) or Brock-Landry-Nadolsky-Yuan (BLNY) parametrizations :
\begin{eqnarray}
F_{NP}^{DWS}&=&\exp\Bigg\{\Big[-\frac{g_1}{2}-g_2\ln(Q/(2Q_{0L}))\Big]b_T^2\Bigg\}\nonumber\\
F_{NP}^{LY}&=&\exp\Bigg\{\Big[-\frac{g_1}{2}-g_2\ln(Q/(2Q_{0L}))\Big]b_T^2 -g_1 g_3\ln(10 x) b_T\Bigg\}\nonumber\\
F_{NP}^{BLNY}&=&\exp\Bigg\{\Big[-\frac{g_1}{2}-g_2\ln(Q/(2Q_{0L}))-g_1 g_3\ln(10 x)\Big]b_T^2\Bigg\}
\label{parametrizations}
\end{eqnarray}
where $g_1$, $g_2$ and $g_3$ are free parameters and $Q_0$ an arbitrary scale.

The resummation approach, extended to non-perturbative regimes using the $b_*$ prescription, can explain the DY data at high energy.
In Ref.~\cite{Landry:2002ix} the authors performed a global fit of FERMILAB, CERN and Tevatron data on $\gamma^*$ and $Z_0$
production using the parametrizations in Eq.~\ref{parametrizations}.
They found that the BLNY parametrization describes data very well, with a $\chi^2/\textrm{dof} \sim1$.
In Ref.~\cite{Konychev:2005iy} Konychev and Nadolsky presented a study on the correlation of the fitted parameter with $b_{max}$.
They found that the data are better described using value of $b_{max}>1 \textrm{ GeV}^{-1}$.
\begin{figure}
\centering
\includegraphics[width=6cm,clip]{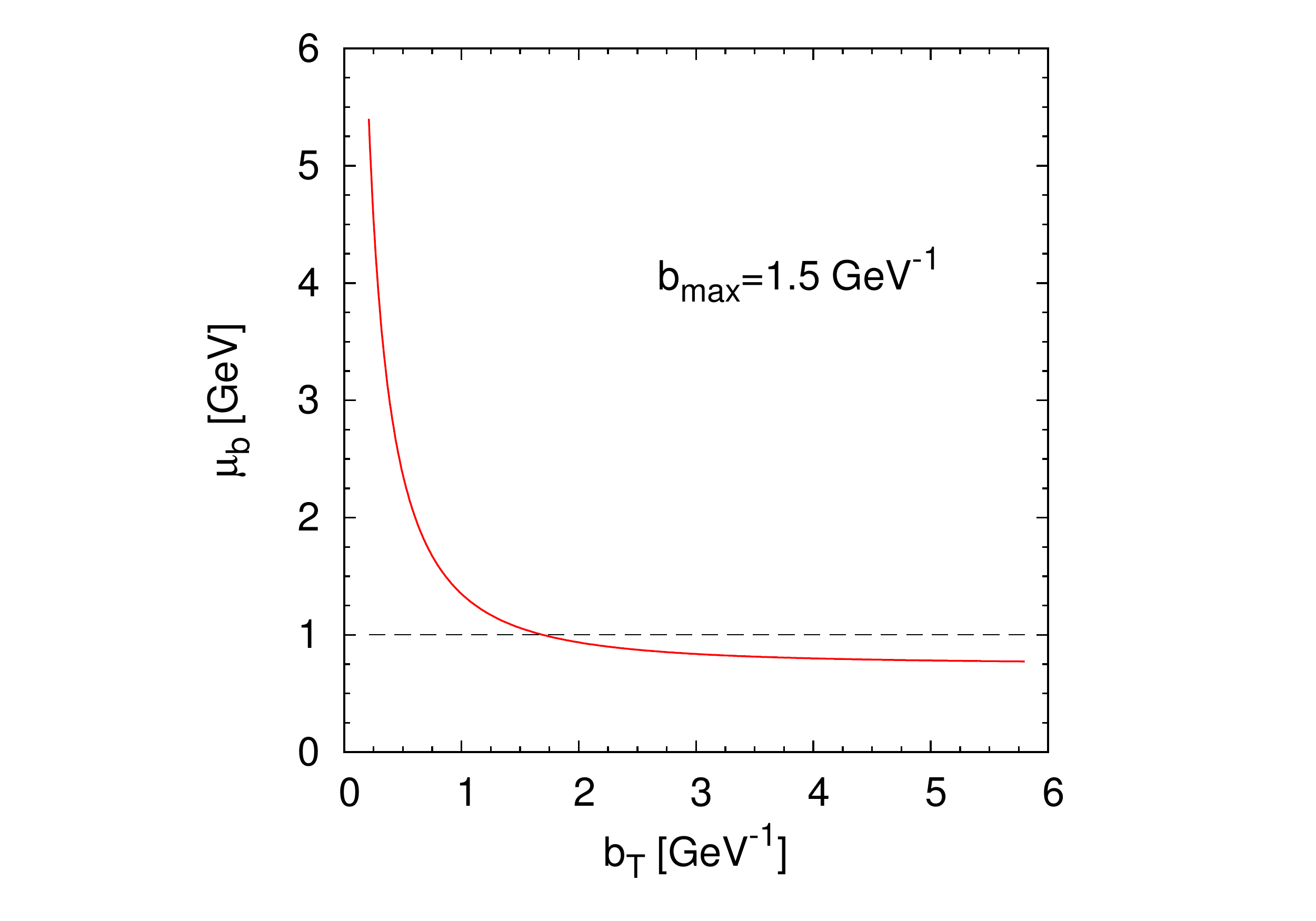}
\caption{$\mu_b$ as function of $b_T$ at fixed value of $b_{max}=1.5\textrm{ GeV}^{-1}$.}
\label{mub_bt}       
\end{figure}
However, as $b_{max}$ becomes larger and larger, $\mu_b$ becomes smaller and smaller and an additional freezing is necessary when $\mu_b$ is too small:
see Fig.~\ref{mub_bt}. Adopting a freezing of $\mu_b$ at 1.3 GeV, they found that $b_{max}=1.5\textrm{ GeV}^{-1}$ is the value which gives the best $\chi^2$.
Moreover they found that substituting the scale $\mu_b$ with $2\mu_b$ in the collinear PDF the description of the data improved.
The treatment of the large $b_T$ region introduces a lot of ``cooking'' prescriptions.
Sometimes in the literature these prescriptions are not fully stated or sufficiently underlined, making it difficult to interpret and reproduce the results.
Additional problems related to this formalism come from the small $b_T$ region.
In fact when $b_T$ goes to zero, $\mu_b$ goes to infinity.
Although theoretically legitimate, from the practical point of
view calculating the PDF at infinity is impossible and again one has to state clearly which strategy is adopted to avoid this problem.
Notice that the $b_*$ prescription is only a prescription, one among others. See for instance the prescriptions adopted in
Refs.~\cite{Ellis:1997sc,Qiu:2000hf,Kulesza:2003wn,D'Alesio:2014vja}.
\section{Unpolarized SIDIS phenomenology}
\label{sidis}
\subsection{Gaussian models}
\label{SubSec:SIDIS-Gauss}
Similarly to the low energy Drell-Yan case, the $P_T$ spectrum of low energy SIDIS experiments is roughly Gaussian.
Here $P_T$ is the transverse momentum of the final produced hadron in the $\gamma^* p$ c.m. frame
measured with respect to the plane containing the initial and final measured leptons. SIDIS processes introduce further difficulties
in order to extract the nucleon parton momentum. In fact, in SIDIS also the fragmentation process contributes to $P_T$.
Therefore, in addition to a Gaussian model in the PDFs, Eq.~\ref{GaussianPDF}, we have to introduce a Gaussian model for the fragmentation process:
\begin{equation}
D_{h/q}(x,p_{\perp};Q^2)=D_{h/q}(z;Q^2)\, \frac{e^{-\frac{p_{\perp}^2}{\langle p_{\perp}^2\rangle}}}{\pi\langle p_{\perp}^2\rangle }
\label{GaussianFF}
\end{equation}
where $p_\perp$ is the transverse momentum of the final produced hadron $h$ w.r.t. the fragmenting parton momentum
and $D_{h/q}(z;Q^2)$ is the ordinary collinear fragmentation function.
Using the Gaussian models in Eqs.~\ref{GaussianPDF} and~\ref{GaussianFF}
we can write the unpolarized SIDIS structure $F_{UU}$ as~\cite{Anselmino:2011ch}:
\begin{equation}
F_{UU}=\sum_q e^2_q \,f_{q/p}(x)\, D_{h/q}(z)\, \frac{\exp(-P_T^2/\langle P_T^2\rangle)}{\pi\langle P_T^2\rangle}
\end{equation}
where
\begin{equation}
 \langle P_T^2\rangle=\langle p_\perp^2\rangle+ z^2 \langle k_\perp^2\rangle\,.
\end{equation}
Using this model the Turin Spin Group~\cite{Anselmino:2013lza} fitted the unpolarized HERMES multiplicity data~\cite{Airapetian:2012ki}
getting a good overall description, $\chi^2/\textrm{dof}=1.69$, and
\begin{eqnarray}
\langle   k_\perp^2\rangle&=&0.57\pm0.08 \textrm{ GeV}^2/c^2\,; \\
\langle p_\perp^2\rangle&=&0.12\pm 0.01\textrm{ GeV}^2/c^2\,.
\end{eqnarray}
A similar analysis of the COMPASS multiplicity data~\cite{Adolph:2013stb} did not give strong conclusive results.
In fact in this case, data do not seem properly normalized. They
exhibit a normalization depending linearly on the $y$ SIDIS variable.
If this normalization correction is taken into account it is possible to describe the data by a Gaussian, although the overall description
is not as good as in the HERMES case;
see Ref.~\cite{Anselmino:2013lza} and the Contribution of O. Gonzalez to these conference proceedings for more details.
The model in Eqs.~\ref{GaussianPDF} and~\ref{GaussianFF} does not take into account any possible flavor dependence,
nor any eventual $x$ dependence in $\langle   k_\perp^2\rangle$ and $z$ dependence in $\langle   p_\perp^2\rangle$.
Since SIDIS measurements are performed by detecting different hadrons in the final state,
one can try to study these further dependencies.
In Ref.~\cite{Signori:2013mda} the authors explored these cases for the HERMES data.
They found a very small improvement using a flavor dependence, getting $\chi^2/\textrm{dof}=1.63\pm0.12$ 
to be compared to $\chi^2/\textrm{dof}=1.72\pm0.11$ for the flavor independent model. 

\subsection{TMD evolution and CSS}
\subsubsection{TMD evolution formalism}
The resummation approach could be applied also to SIDIS data. However current phenomenological analyses of SIDIS data refer more generally
to the TMD evolution scheme as presented by J.~Collins in Ref.~\cite{Collins:2011zzd}. In fact, we have to remember that, even if related to transverse momentum,
the CSS scheme does not define any TMD. It is the collinear cross section, as a whole, that is resummed.
Nevertheless the CSS is valid also in the context of the Collins's TMD evolution and the author himself defines the TMD evolution exposed in Ref.~\cite{Collins:2011zzd}
the \textit{new CSS} as it was a simple extension of the CSS.

For instance in the new TMD CSS formalism the DY cross section simply reads:
\begin{equation}
\frac{d\sigma}{dQ^2dy d q_T^2}=H^2(Q)\int\frac{d^2 \boldsymbol{b}_T e^{i \boldsymbol{q}_T\cdot\boldsymbol{b_T}}}{(2\pi)^{2}}\Bigg\{
\sum_{q,\bar{q}}e^2_q\tilde{F}_{1q}(x_1,b_T,Q,\zeta_{F1})\tilde{\bar{F}}_{2q}(x_2,b_T,Q,\zeta_{F2})\Bigg\} + Y(x_1,x_2,P_T,Q)\label{Master_RAC1}
\end{equation}
where $H^2(Q^2)$ is a process dependent hard factor~\cite{Aybat:2011vb} and the
TMD PDF  $\tilde{F}_{q}(x,b_T,Q,\zeta_F)$ is:
\begin{eqnarray}
&&\tilde{F}(x,b_T,Q,\zeta_{F})=\left({\frac{\sqrt{\zeta_F}}{\mu_b}}\right)^{\tilde{K}(b_*,\mu_b)}
\sum_j\int_x^1 \frac{d y}{y} \tilde{C}_{f/j}(x/y, b_*,\mu_b,\mu_b^2) f_j(y,\mu_b)\nonumber\\
&&\exp\left\{\int_{\mu_b}^Q \frac{d \mup}{\mup} \gamma_F(\mup;1)-\ln\left(\frac{\sqrt{\zeta_F}}{\mup}\right)\gamma_K(\mup)\right\}
\exp\left\{-g_{P}(x,b_T)-g_K(b_T)\ln\left(\frac{\sqrt{\zeta_F}}{\sqrt{\zeta_{F0}}}\right)\right\}\label{Master_RAC4}\,.
\end{eqnarray}

$\tilde{F} (x,b_T,Q,\zeta_{F})$ depends as usual on $x$, $b_T$ and $Q$ and on the new variable $\zeta_F$.
As explained extensively in Ref.~\cite{Collins:2011zzd}, $\zeta_F$ is related to the rapidity divergences that appear in the TMD soft factors.
In SIDIS and DY, where the factorization is proven, we have always a product of two TMDs.
It is possible to show that the product of the corresponding $\zeta$ is such that $\zeta_1*\zeta_2=Q^4$.
Therefore, as a practical procedure, one can simply set $\zeta\equiv Q^2$.
$\tilde{K}$ is related to the evolution of the TMD w.r.t. $\zeta_F$:
\begin{equation}
\frac{\partial\ln\tilde{F}(x,b_T,\mu,\zeta_{F}) }{\partial \ln \sqrt{\zeta_F}}= \tilde{K}(b_T,\mu)\,.
\end{equation}
$\tilde{C}_{f/j}$ are the process independent Wilson coefficients; 
$f_j(x,Q)$ is the standard collinear PDF; $\gamma_F$ and $\gamma_K$
are anomalous dimensions of $\tilde{F}(x,b_T,Q,\zeta_{F})$ and $\tilde{K}(b_T,\mu)$.
They are calculable in QCD:
\begin{equation}
 \frac{d\tilde{K}(b_T,\mu) }{d\ln\mu}=-\gamma_K(\mu)\,;
\end{equation}
\begin{equation}
 \frac{d\ln\tilde{F}(x,b_T,\mu,\zeta_{F}) }{d\ln\mu}=\gamma_F(\mu;\zeta_F/\mu^2)\,. 
\end{equation}
$g_P(x,b_T)$ and $g_K(b_T)$ are two non-perturbative functions.
$g_P(x,b_T)$ depends on the particular hadron considered while $g_K(b_T)$
is universal: it is the same for any TMD.
$b_*$ is defined as in Eq.~\ref{bstar}.

Eqs.~\ref{Master_RAC1} and~\ref{Master_RAC4} appear rather different from Eqs.~\ref{MasterCSS} and~\ref{W2}.
However, this is not really true.
First of all, notice that the convolution of the Wilson coefficients and the PDF, $C\otimes f$,
is present in both the CSS and the new TMD CSS.
The main difference is that in the CSS scheme the Wilson coefficients are process dependent.
To avoid that, in the new TMD CSS all the parts of the Wilson coefficient
that in the old CSS were process dependent are put in the hard factor $H(Q^2)$~\cite{Aybat:2011vb}.
Apart from that, provided $\zeta_F\equiv Q^2$, the process independent part of the old and new CSS $C$'s are the same.
Moreover setting $b_*$ and $\mu_b$ as in Eq.~\ref{mub_bt_replace} then $\tilde{K}(b_*\mu_b)=0$.
Finally, at least at fixed order in $\alpha_s$ one can explicitly show that the Sudakov factor $S^{CSS}$, defined by Eq.~\ref{S}, is given by:
\begin{equation}
 S^{CSS}(b_T,Q)=2\int_{\mu_b}^Q \frac{d \mup}{\mup} \gamma_F(\mup;Q^2/\mup^2) 
\end{equation}
provided again that $\zeta_F\equiv Q^2$ and $b_*$ and $\mu_b$ are defined as in Eq.~\ref{mub_bt_replace}.
Therefore:
\begin{equation}
\tilde{F} (x,b_T,Q,\zeta_{F}\equiv Q^2)= \sum_j \tilde{C}_{f/j}\otimes f_j(x,\mu_b)\exp\left\{\frac{1}{2}S^{CSS}(b_*,\mu_b)\right\}
\exp\left\{-g_{P}(x,b_T)-g_K(b_T)\ln\left(\frac{Q}{Q_0}\right)\right\}
\end{equation}
where the correspondence with the CSS formalism is more evident. Moreover, notice that all the parametrizations in Eq.~\ref{parametrizations}
are of the type $\exp\left\{-g_{P}(x,b_T)-g_K(b_T)\ln\left(\frac{Q}{Q_0}\right)\right\}$\,.

In literature Eq.~\ref{Master_RAC4} is not always used in this form.
In fact, it is sometimes more convenient to build the ratio between two TMDs at two different scales:
\begin{eqnarray}
\frac{\tilde{F}(x,b_T,Q,Q^2)}{\tilde{F}(x,b_T,Q_0, Q_0^2)}&=&
\left(\frac{Q}{Q_0}\right)^{\left[-\int_{\mu_b}^{Q_0}\frac{d\mup}{\mup}\gamma_K(\mup)\right]}
\exp\left\{\int_{Q}^{Q_0}\frac{d\mup}{\mup}\left[\gamma_F(\mup;1)-\gamma_K(\mup)\ln\left(Q/\mup\right)\right]\right\}
\exp\left[-g_K(b_T) \ln\left(Q/Q_0\right)\right]\nonumber\\
&=&\tilde{R}(Q,Q_0,b_T)\exp\left[-g_K(b_T) \ln\left(Q/Q_0\right)\right]\label{RAP-master2}
\end{eqnarray}
and define a TMD at scale $Q$ as function of the same TMD at scale $Q_0$:
\begin{eqnarray}
\tilde{F}(x,b_T,Q,Q^2)=\tilde{F}(x,b_T,Q_0, Q_0^2)\,\tilde{R}(Q,Q_0,b_T)\exp\left[-g_K(b_T) \ln\left(Q/Q_0\right)\right]\,.
 \label{TMDIO}
\end{eqnarray}
In this review I will call this evolution equation ``Input-Output TMD Evolution'' (TMD IO).
Notice that the ratio in Eq.~\ref{RAP-master2} is the same for the unpolarized TMD and the first derivative of the Sivers function~\cite{Aybat:2011ge}:
\begin{equation}
 \frac{\tilde{f}_{1T}^{\prime\perp}(x,b_T,Q,\zeta_{F})}{\tilde{f}_{1T}^{\prime\perp}(x,b_T,Q_0,\zeta_{F0})}
\equiv \frac{\tilde{F}(x,b_T,Q,\zeta_{F})}{\tilde{F}(x,b_T,Q_0,\zeta_{F0})}\,.
\label{ratiosivers}
 \end{equation}
 The unpolarized fragmentation functions follow similar evolution equations, see Refs.~\cite{Collins:2011zzd,Aybat:2011zv}
 \subsubsection{TMD/CSS phenomenology}
 In Ref.~\cite{Echevarria:2014xaa} an application of Eqs.~\ref{Master_RAC1} and~\ref{Master_RAC4} to a global fit of DY and SIDIS data is presented.
 However, the authors use one important approximation: the Wilson coefficients $C$ in Eq.~\ref{Master_RAC4}
 are calculated at tree-level and the collinear PDFs are computed at leading order (LO),
 while the coefficient of the anomalous dimensions are calculated consistently at Next to Leading Log (NLL).
 For the non-perturbative model they use a model \textit{a la} DWS, see Eq.~\ref{parametrizations},
 which may work less well than the BLNY model~\cite{Landry:2002ix}, at least for DY.
 The description they obtain is in fair agreement with low and high energy DY data.
 However SIDIS data are only qualitatively reproduced. In particular, HERMES data for $\pi^+$ production are poorly described.
 No $\chi^2$ of the fit is provided, so it is impossible to make quantitative statements.
 
 In Ref.~\cite{Aybat:2011ta} Aybat, Rogers and Prokudin (ARP) applied, for the first time, the TMD evolution
 to the study of the Sivers function using Eq.~\ref{TMDIO}.
 As a product they have also a model for the unpolarized cross section. As input function they use a simple Gaussian model
 \begin{equation}
 \tilde{F}(x,b_T,Q_0, Q_0^2)=f_{q/p}(x, Q_0)\exp\left[-\frac{\langle k_{\perp}^2\rangle}{4} b_T^2\right]
\label{GaussRAP}
 \end{equation}
with $ \langle k_{\perp}^2\rangle$ extracted form Ref.~\cite{Anselmino:2005nn}.
The parameters $b_{max}$ and $g_2$ are those of the BLNY parametrization in Ref.~\cite{Landry:2002ix}.
In Ref.~\cite{Sun:2013dya} Sun and Yuan (SY) point out that if one uses Eq.~\ref{RAP-master2} with the Gaussian model of Eq.~\ref{GaussRAP} as an input,
and the parameters in Ref.~\cite{Aybat:2011ta},
then the description of the unpolarized DY data badly fails, see Fig. 1 of Ref.~\cite{Sun:2013dya}.
Sun and Yuan, still using Eq.~\ref{RAP-master2}, propose to modify the factor $R^{\prime}=R(Q,Q_0, b_T)\exp(g_K(b_T)\ln(Q/Q_0))$ in the following way:
\begin{equation}
R^{\prime}=2 C_F\int_{Q_0}^{Q}\frac{d\mup}{\mup}\frac{\alpha_s(\mup)}{\pi}\left[\ln\left(\frac{Q^2}{\mup^2}\right)+\ln\frac{Q_0^2 b^2}{C_1^2}-\frac{3}{2}\right]\,.
\label{YuanSuSudakov}
\end{equation}

This approximation, rather difficult to justify (see J.~Collins contribution to this conference), works approximately for low energy DY data while
it cannot be used for high energy DY data. Notice that the new evolutor does not depend on $\mu_b$.
As an input function the authors used again a model like Eq.~\ref{GaussRAP}
but with the parameters of Ref.~\cite{Schweitzer:2010tt}, setting the initial scale, $Q_0$, as
the average $Q$ of the HERMES experiment. However, despite these modifications, the description of the HERMES experiment
(which is the input for the evolution) is very bad, see Fig.1 of Ref.~\cite{Sun:2013dya}.
The description of COMPASS data seems to be, instead, satisfactory. However it is worth to say that they considered only some particular bins of COMPASS,
allowing for a normalization factor.

Notice that Eq.~\ref{RAP-master2} and Eq.~\ref{Master_RAC4} are formally equivalent. The fact that Eq.~\ref{RAP-master2}
failed the description of DY data is due to the Gaussian approximation of the input function.
For instance in the Gaussian approximation the $x$ and $b_T$
dependences are completely factorized, while Eq.~\ref{Master_RAC4} clearly shows that this is impossible.
Similarly, if the input function is a Gaussian some of the $b_T$ dependences are different.
Eq.~\ref{RAP-master2} is suitable only for an approximate study of the Sivers asymmetry
(this in fact was the original use of Eq.~\ref{RAP-master2} in Ref.~\cite{Aybat:2011ta})
where we are not sensitive to much detail since the observable is itself a ratio.
\section{Sivers effect}
\label{Sec:Sivers}
The Sivers asymmetry plays a crucial role in the investigation of TMD evolution. In fact the Sivers asymmetry
is the simplest TMD asymmetry that we can study since it involves only
the Sivers function and the unpolarized TMDs. Moreover, as mentioned already in the introduction, the Sivers function exhibits a non trivial universality,
changing sign in SIDIS and DY. The Sivers function $f_{1T}^{\perp}(x,k_{\perp})$ is a TMD of rank 1, \textit{i.e.}
the first non-zero moment of the function is its first moment. For this reason the object which evolves is not the Sivers function
but the first derivative of the Sivers function in the $b_T$ space.  
The TMD CSS evolution of the first derivative of the Sivers function, $\tilde{f}_{1T}^{\prime\perp}(x,b_T,Q,\zeta_{F})$, is the following:
\begin{eqnarray}
&&\tilde{f}_{1T}^{\prime\perp}(x,b_T,Q,\zeta_{F})= \left({\frac{\sqrt{\zeta_F}}{\mu_b}}\right)^{\tilde{K}(b_*,\mu_b)}
\frac{m_p b_T}{2}\sum_j\int_{x_1}^1 \int^1_{x_2}\frac{d x_1 d x_2}{x_1 x_2}\Bigg\{
\tilde{C}_{f/j}^{Siv}(x_1,x_2, b_*,\mu_b,\mu_b^2) \,T_{Fj}(x_1,x_2,\mu_b)\Bigg\}\nonumber\\
&&\exp\left\{\int_{\mu_b}^Q \frac{d \mup}{\mup} \gamma_F(\mup;1)-\ln\left(\frac{\sqrt{\zeta_F}}{\mup}\right)\gamma_K(\mup)\right\}
\exp\left\{-g_{P}^{Siv}(x,b_T)-g_K(b_T)\ln\left(\frac{\sqrt{\zeta_F}}{\sqrt{\zeta_{F0}}}\right)\right\}\label{Master_RAC}
\end{eqnarray}
where $\tilde{C}_{f/j}^{Siv}$ are the Wilson coefficients for the Sivers function;
$T_{Fj}(x_1,x_2,\mu)$ is the Qiu-Sterman functions, which is the collinear ``equivalent'' of the Sivers function;
$g_P^{Siv}(x,b_T)$ is a non-perturbative function that similarly to $g_P(x,b_T)$ depends on the hadron considered;
all the other quantities are the same appearing in Eq.~\ref{Master_RAC4}.
As we can see comparing Eq.~\ref{Master_RAC4} and Eq.~\ref{Master_RAC},
the evolutor of the first derivative of the Sivers function is the same of the unpolarized function, from which we can get Eq.~\ref{ratiosivers}.
As mentioned in the previous section, the first paper devoted to the TMD evolution of the Sivers function was Ref.~\cite{Aybat:2011ta}.
Using as input function the Sivers function extracted from Ref.~\cite{Anselmino:2008sga},
Aybat, Rogers and Prokudin showed that HERMES and COMPASS data
on the Sivers effect were compatible to each other.
A first global fit of HERMES and COMPASS data on the Sivers asymmetry
taking into account the TMD CSS evolution was performed in Ref.~\cite{Anselmino:2012aa} by Anselmino, Boglione and Melis (ABM).
The authors used the TMD IO version of the TMD evolution, Eq.~\ref{RAP-master2}.
The input function was Gaussian similarly to Ref.~\cite{Aybat:2011ta}.
The parameter $g_2$ was that extracted in Ref.~\cite{Landry:2002ix} while the Sivers function parameters are fitted.
The fit gave a good description of the data, corresponding to a $\chi^2/\textrm{dof}=1.02$.
In order to compare this extraction with the standard approach of Ref.~\cite{Anselmino:2008sga},
ABM also performed a traditional Gaussian fit, without TMD evolution,
as those illustrated in the Sects.~\ref{SubSec:Drell-Yan-Gauss} and~\ref{SubSec:SIDIS-Gauss}.
In this case the  $\chi^2/\textrm{dof}$ was 1.26, denoting a slightly worse description w.r.t. the TMD evolution scheme although still overall good.
The study of the Sivers asymmetry presented in Ref.~\cite{Sun:2013dya} by SY using the modified Sudakov, Eq.~\ref{YuanSuSudakov},
led to similar $\chi^2$: $\chi^2/\textrm{dof}=1.08$.
All the mentioned papers used a IO version of the TMD CSS evolution equation.
In Ref.~\cite{Echevarria:2014xaa} Echevarria, Idilbi, Kang and Vitev (EIKV) used for the first time Eq.~\ref{Master_RAC}
using however, as in the unpolarized case, the Wilson coefficient at tree-level. The $T_F$ was approximated as the first moment of the Sivers function
and parameterised as in Ref.~\cite{Anselmino:2008sga}. They got an overall good description of the data, with a $\chi^2/\textrm{dof}=1.3$.
These results are summarized in table~\ref{tableSiv}.
\begin{table}
\begin{tabular}{llll}
\hline
Sivers Effect&&\\
&&\\
\hline
Group                             & Model                & FIT & $\chi^2/\textrm{dof}$  \\
\hline
ARP~\cite{Aybat:2011ta}           & TMD IO               & NO  & Qualitatively OK \\
ABM ~\cite{Anselmino:2012aa}      & Gaussian             & YES & 1.26 \\
ABM~\cite{Anselmino:2012aa}       & TMD IO               & YES & 1.02 \\
SY~\cite{Sun:2013dya}             &TMD IO+ Mod. Sudakov  & YES & 1.08 \\
EIKV ~\cite{Echevarria:2014xaa}   &TMD CSS+ C at LO      & YES & 1.3 \\
\hline
\end{tabular}
\caption{Comparison among different Sivers Asymmetry fits in SIDIS. Notice that each group may have considered slightly different sets of data.
Here we want to show that description of the Sivers asymmetry is roughly equivalent using different models.}
\label{tableSiv}       
\end{table}
\section{Conclusions}
In the last years a lot of progress has been achieved in the study of the TMDs. The HERMES, COMPASS and JLAB Collaborations, the high energy facilities Tevatron and LHC,
the $e^+e^-$ facilities BELLE and BABAR, have provided a huge quantity of new experimental data related to the TMDs.
Theoretically, Ref.~\cite{Collins:2011zzd} opened the exploration of the TMD evolution,
initiating a big theoretical debate on its meaning, possible improvements and alternatives.
The picture that emerges is that TMDs are strongly related to their ``cousins'',
the collinear PDFs and the collinear correlators. Somehow the TMD, in the region where $\Lambda_{\textrm{QCD}}\ll k_{\perp}\ll Q$,
is just a resummed collinear object: an unpolarized PDF for the unpolarized TMD, a Qiu-Sterman function for the Sivers function, etc..
However in the region of small transverse momenta, the perturbative approach fails and we have to introduce non-perturbative functions and parameters.
This region, where everything is non-perturbative, is the region where the original idea of TMD was born.
The Gaussian models, widely used in the literature, are the simplest examples of non-perturbative functions.
To get the full picture it is clear that we need to explore the whole region of the transverse momentum spectra
in different experiments. Unfortunately the data are, in this context, sparse.
The studies on TMDs in the last years were mainly devoted to asymmetries, like the Sivers or the Collins asymmetries.
However, if we want to be quantitative and make reliable predictions, we have to change our mind and focus on the unpolarized processes.
It is worth to mention that unpolarized TMDs enter in (the denominator of) any asymmetry. From Tab.~\ref{tableSiv},
which summarizes the results of Sect.~\ref{Sec:Sivers}, we can see that different groups are able
to explain the Sivers effect,
using very different approaches. Even a simple Gaussian model gives good results.
It is clear that the extracted Sivers functions are really different from each other (quantitatively).
In Tab.~\ref{tableUnp} we show, in a schematic form, how these groups describe the corresponding unpolarized data with their models.
The situation is definitively not satisfactory and suggests to take the results on the Sivers asymmetry (and on all the TMDs) with a grain of salt.
Future experiments like the EIC~\cite{Accardi:2012qut} or the COMPASS DY program~\cite{Gautheron:2010wva} will certainly help to shed light on the field.
\begin{table}
\begin{tabular}{lllll}
\hline
\hline
Group                             & Model                & HERMES & DY Low $\sqrt{s}$  & DY High $\sqrt{s}$ \\
\hline
ARP~\cite{Aybat:2011ta}           & TMD IO               & NO    & NO             & NO \\
ABM ~\cite{Anselmino:2012aa}      & Gaussian             & YES   & YES Separately & NO\\
ABM~\cite{Anselmino:2012aa}       & TMD IO               & NO    & NO             & NO \\
SY~\cite{Sun:2013dya}             &TMD IO+ Mod. Sudakov  & NO    & YES      & NO \\
EIKV ~\cite{Echevarria:2014xaa}   &TMD CSS+ C at LO      & NO    & YES & YES\\
\hline
\end{tabular}
\caption{Description of unpolarized data by the same model in Tab.~\ref{tableSiv}. 
COMPASS data are not considered here since each paper considered very different/partial/incomplete COMPASS bins.}
\label{tableUnp}      
\end{table}




%
\bibliography{stefano_melis_bibtex}
%
%
%

\end{document}